Extending earlier work on a vibronic theory of the $F_A$ centers in alkali halides, the reorientation is now considered of an off-centered $Li^+$ impurity, either isolated or one near an F center. We derive analytically periodic potential energy barriers between metastable reorientational off-center sites, the barriers hindering the impurity rotation around the normal lattice site. Applying to a specific model, electron-mode coupling constants are calculated up to third order of the expansion of the coupling energy in the $T_{1u}$ vibrational mode coordinates. The third-order coupling brings about additional renormalization of the effective vibrational frequency controlling the reorientation.

1. Introduction

A great deal of experimental work has accumulated over the years on the optical, electric, and magnetic properties of small-size impurities in alkali halides, either isolated or near an F center. Although the theory has followed suit in some cases it has lagged behind in others and there seems to be no consensus on just what happens as the impurity is placed in a polarizable crystalline environment. For a review of earlier experimental and theoretical work, see [1],[2]. Recently we analyzed experimental and numerical literature data to show that a vibronic approach to the impurity-lattice problem was reasonably feasible [3]. Among other things, we confirmed earlier findings [4],[5] that under well-defined conditions, controlled mainly by the electron-phonon coupling (mixing) strength, the impurity would go off center and derived simple expressions for the off-centered displacement, both for an isolated impurity and for one near an F center. In particular, the prototype $Li^+$ impurity and the F center were shown to undergo a dipole-dipole coupling which led to shrinking the off-center radius of the $Li^+$ ion when near an F center. Numerical estimates of F center and off-centered impurity polarizabilities led to radial displacements in concert with experimental and computerized data. From the vibronic viewpoint, the enigmatic behavior of the far-infrared absorption bands attributed to a $Li^+$ impurity was also understood. (See [3] and the references therein for a brief survey of the topic.)

The present work in four parts is aimed at extending our vibronic study along similar lines. Whereas the previous emphasis has laid on the off-center displacement and some of its immediate implications, this one centers on the rotational behavior of an impurity already off-centered. In part I, we confirm analytically earlier expectations that off-center displacements and potential-energy barriers hindering the free rotation of an off centered species around a central site both result from the mixing by the same vibrational $T_{1u}$ mode of a pair of electronic states of an impurity-lattice cluster. In part II, the rotational problem of an off-centered entity is considered in greater detail relevant to an $F_A$ center, due to its inherent 2-D character. $Li^+$ off-center polarizabilities relevant to specific experimental situations are also discussed. In part III, our analytical results are compared with numerical data by extended Hückel calculations. Finally, relaxation rates are derived in part IV of impurity reorientation near an $F_A$ center.

## 2. Hamiltonian background

We consider an impurity ion embedded in a crystalline medium. The latter is regarded as a system of lattice oscillators, that is, vibrating ion cores, as well as outer-shell electrons coupled to the oscillators. The relevant Hamiltonian builds up by electron, lattice and electron-lattice terms, respectively:

$$H = H_e + H_L + H_{eL} \tag{1}$$

$H_e$ is the static electronic Hamiltonian at a fixed lattice when all the oscillators are frozen in unperturbed equilibrium positions at $Q^\sim = 0$ ($Q^\sim$ is the domain of nuclear coordinates):

$$H_e = \sum [p_e^2 / 2m_e + V_e(r_e, 0^\sim)] = \sum_\tau E_\tau a_\tau^\dagger a_\tau \tag{2}$$

the sum being over the coordinates $r_e$ and momenta $p_e$ of all the electrons, $m_e$ are the electron masses. $V_e(r_e, 0^\sim)$ is the static potential which the electrons "see" when the nuclei are at rest. When they are not, the electronic potential varies following parametrically the nuclear motion (adiabatic approximation). In second quantization terms, $E_\tau$ are one-electron energies, $a_\tau^\dagger$ ($a_\tau$) are electron creation (annihilation) operators, and the sum is over the electronic states $\tau$. The modulated potential $V_e(r_e, Q^\sim)$ can be expanded into a power series in $Q^\sim$ to give

$$V_{eL}(r_e, Q^\sim) = V_e(r_e, Q^\sim) - V_e(r_e, 0^\sim) =$$

$$\sum_i b_i(r_e) Q_i + \sum_{ij} c_{ij}(r_e) Q_i Q_j + \sum_{ijk} d_{ijk}(r_e) Q_i Q_j Q_k + ... \tag{3}$$

The mixed electron-lattice terms in (3) effect electron coupling to the lattice oscillators. Accordingly we construct an electron-lattice interaction Hamiltonian of the form

$$H_{eL} = \sum_{\alpha\beta} V_{eL}(Q^\sim)_{\alpha\beta} a_\beta^\dagger a_\alpha \tag{4}$$

where the subscripts αβ attached to equation (3) terms run over the eigenstates of $H_e$: $|\alpha,\beta> \equiv |\alpha,\beta,0\tilde{\;}>$, to be specified as $|\alpha> = |a_{1g}>$ and $|\beta_i> = |t_{1ui}>$ with i = x,y,z [3], etc. Generally, Hamiltonian (4) contains both band-diagonal α = β and band-off-diagonal α ≠ β terms.

In the absence of an electron-lattice interaction, the lattice Hamiltonian reads

$$H_L = \tfrac{1}{2} \sum_s [P_s^2 / M_s + M_s \omega_s^2 Q_s^2] + \tfrac{1}{2} \sum_l [P_l^2 / M_l + M_l \omega_l^2 Q_l^2] + ... \quad (5)$$

Here $Q_s$, $P_s$, $\omega_s$, and $M_s$ stand for the coordinates, momenta, angular frequencies, and reduced masses, respectively, of the lattice oscillators. The coordinates, etc. of the reorientation-promoting mode $Q_l$, $P_l$, $\omega_l$, and $M_l$ are taken out of the main sum. The dots imply mixed coupling terms of the form $g\sum_{sl}Q_sQ_l$ as well as higher-order terms. By second quantization

$$H_L = \sum_s \hbar\omega_s b_s^\dagger b_s + \sum_l \hbar\omega_l b_l^\dagger b_l + ...$$

Where $b_l^\dagger$ ($b_l$) are the phonon creation (annihilation) operators; however, with the adiabatic approximation to be used throughout, the vibrational coordinates Q will be regarded as c-numbers.

We next make several assumptions simplifying the mathematical problem without sacrificing the basic physics:

(i) Harmonic approximation: omitting the dots in (5) to regard $H_L$ diagonalized with respect to the nuclear coordinates.
(ii) Predominating promoting-mode coupling: the electron-lattice coupling with the promoting mode $Q_l$ prevails, so as to discard the dots in (5). This implies that the $gQ_sQ_l$ lattice modes−promoting mode coupling terms are too small to affect the energy balance; this holds better at low temperatures.
(iii) Adopting a coupling scheme confined to the linear, second-order and third-order terms thereby neglecting the dots in (3).

Discarding the lattice--mode coupling under (i), (ii) makes considering the sum terms in (5) unimportant and (1) reduces to

$$H = \tfrac{1}{2} \sum_\tau \sum_i [P_i^2 / M_i + M_i \omega_i^2 Q_i^2]_\tau + \sum_\tau E_\tau a_\tau^\dagger a_\tau +$$

$$\sum_{\alpha\beta} \sum_i [b_i(r_e) Q_i + \sum_j c_{ij}(r_e) Q_i Q_j + \sum_{jk} d_{ijk}(r_e) Q_i Q_j Q_k]_{\alpha\beta} a_\beta^\dagger a_\alpha$$

To tackle the impurity-lattice problem by Born & Oppenheimer's theorem we define an adiabatic Hamiltonian $H_{AD} \equiv H - \sum_l [P_l^2/2M_l]_\tau$ thereby discarding the nuclear kinetic energy operator:

$$H_{AD} = \tfrac{1}{2} \sum_\tau \sum_i [M_i \omega_i^2 Q_i^2]_\tau + \sum_\tau E_\tau a_\tau^\dagger a_\tau +$$

$\sum_{\alpha\beta} \sum_i [\ b_i\ (r_e)\ Q_i + \sum_j c_{ij}(r_e)\ Q_i\ Q_j + \sum_{jk} d_{ijk}\ (r_e)\ Q_i\ Q_j\ Q_k\ ]_{\alpha\beta}\ a_\beta^\dagger a_\alpha$

Once eigenvalues $E_{U/L}(Q_i)$ and eigenstates $\Phi(Q_i)$ of the adiabatic Hamiltonian $H_{AD}$ have been deduced, solving for the vibronic problem

$$\{\sum_l [\ P_l^2\ /\ 2M_l\ ]_\tau + E_{U/L}(Q_i)\} u(Q_i) = E_{vib} u(Q_i) \qquad (8)$$

would yield rotational states and eigenenergies for the impurity cluster.

### 3. Off-center impurity in alkali halide

#### 3.1. General arguments

There is an accumulating evidence that substitutional impurities in ionic crystals displace off-center as a result of the vibronic mixing of nearly-degenerate electronic states at the impurity ion by ungerade vibrations which render the normal lattice site configuration unstable. For the well known example of a $Li^+$ ion in an alkali halide crystal, we assume the availability of $a_{1g}$- and $t_{1u}$-symmetry impurity cluster states to be mixed by an odd-parity local vibrational mode of symmetry $T_{1u}$. We next retain the band-off-diagonal terms $\alpha \neq \beta$ to construct an electron-lattice mixing Hamiltonian of the form

$H_{eL} = \sum_i [V_{eL}]_{\alpha\beta i}\ (a_{\beta i}^\dagger\ a_\alpha + a_\alpha^\dagger\ a_{\beta i}) \equiv$

$\sum_i [\ b_i\ (\mathbf{r_e})Q_i + \sum_j c_{ij}(\mathbf{r_e})\ Q_i\ Q_j + \sum_{jk} d_{ijk}(\mathbf{r_e})\ Q_i\ Q_j\ Q_k]_{\alpha\beta i}\ (a_{\beta i}^\dagger\ a_\alpha + a_\alpha^\dagger\ a_{\beta i})$

where $\alpha = a_{1g}$ and $\beta_i = t_{1ui}$ ($i = x,y,z$). The eigenvalues and eigenstates of the adiabatic Hamiltonian $H_{AD}$ with $H_{eL}$ from (3) and (4) at $c_{ij} = d_{ijk} = 0$ have been derived and discussed elsewhere [3].

We next extend these results by solving for Schrödinger's equation

$H_{AD} | n, Q_l \rangle = E_{AD} | n, Q_l \rangle$

by means of the linear combination

$$| n, Q_l \rangle = A(Q_l)\ |\ a_{1g}\rangle + \sum_i A_{\beta i}(Q_l)\ |\ t_{1ui}\rangle \qquad (10)$$

where

$H_{AD} = \tfrac{1}{2} \sum_\tau \sum_i [M_i\ \omega_i^2\ Q_i^2]_\tau +$

$\sum_\tau E_\tau a_\tau^\dagger a_\tau + \sum_{\alpha\beta} \sum_i [\ b_i\ (r_e)\ Q_i + \sum_j c_{ij}(r_e)\ Q_i\ Q_j + \sum_{jk} d_{ijk}\ (r_e)\ Q_i\ Q_j\ Q_k\ ]_{\alpha\beta}\ a_\beta^\dagger a_\alpha \qquad (11)$

We set $c_{ij} = 0$, since $\alpha$ and $\beta_i$ are opposite-parity states because Hamiltonian (1) should conserve parity, to get a coupled system of equations for the amplitudes:

$[H_{AD\alpha\alpha} - E]A_\alpha + H_{AD\alpha x} A_x + H_{AD\alpha y}A_y + H_{AD\alpha z}A_z = 0$

$H_{ADx\alpha} A_\alpha + [H_{ADxx} - E]A_x + H_{ADxy} A_y + H_{ADxz} A_z = 0$

$H_{ADy\alpha} A_\alpha + H_{ADyx} A_x + [H_{ADyy} - E]A_y + H_{ADyz} A_z = 0$

$H_{ADz\alpha} A_\alpha + H_{ADzx} A_x + H_{ADzy} A_y + [H_{ADzz} - E]A_z = 0$

where

$H_{AD\,\delta\mu} = \{\sum_\tau E_\tau + \frac{1}{2} \sum_\tau \sum_l [(M_l \omega_l)^2 Q_l^2]_\tau \}\delta_{\delta\tau}\delta_{\tau\mu} +$

$\sum_i [b_i Q_i + \sum_j c_{ij} Q_i Q_j + \sum_{jk} d_{ijk} Q_i Q_j Q_{k\,\alpha\beta i}](\delta_{\delta\alpha}\delta_{\beta i\mu} + \delta_{\delta\beta i} \delta_{\alpha\mu})$

with $\delta, \mu = \alpha, \beta_i$ and $E_{\alpha,\beta i} = \pm\frac{1}{2}E_{\alpha\beta}$. We solve for the energy:

$$E_{U/L}(Q_i) = \frac{1}{2} \sum_i K_i Q_i^2 \pm [\sum_i (b_i Q_i + \sum_{jk} d_{ijk} Q_i Q_j Q_k)^2 + (E_{\alpha\beta}/2)^2]^{1/2} \quad (12)$$

with $K_i = M_i \omega_i^2$. Using (12) the vibronic Hamiltonian in (8) is

$H_{vib} = - (\hbar^2 / 2M) \sum_i (\partial^2 / \partial Q_i^2) + \frac{1}{2} \sum_i K_i Q_i^2 \pm$

$$[\sum_i (b_i Q_i + \sum_{jk} d_{ijk} Q_i Q_j Q_k)^2 + (E_{\alpha\beta} / 2)^2]^{1/2} \quad (13)$$

Now while the adiabatic energy surface is mainly controlled by $b_i$, the mixed $b_i d_{ijk}$ terms produce sites for hindered rotation [4].

### 3.2. Specific model

To illustrate the foregoing statement we choose a $T_{1u}$-symmetry d-tensor: side diagonal $d_{ijj} = d_b$, main diagonal $d_{iii} = d_c$, and $d_{ijk} = 0$ otherwise [5]. The first-order tensor $b_i$ is also reduced to a single component and so is the spring constant: $b_i = b$, $M_i \omega_i^2 = M\omega^2$. Introducing a radial' coordinate $Q = \sqrt{(\sum_i Q_i^2)}$ and neglecting the small terms under the root sixth order in $Q_i$ we get:

$$E_{U/L} (Q_i) = \frac{1}{2}M\omega^2 Q^2 \pm \{(bQ)^2 + 2b[(d_c - d_b) \sum_i Q_i^4 + d_b Q^4] + (E_{\alpha\beta} / 2)^2\}^{1/2} \quad (14)$$

making use of $Q^4 = \sum_i Q_i^4 + 2(Q_x^2 Q_y^2 + Q_y^2 Q_z^2 + Q_z^2 Q_x^2)$.

Neglecting for a while all d-terms we get

$$E_{U/L} (Q)_{d=0} = \frac{1}{2} M\omega^2 Q^2 \pm \{(bQ)^2 + (E_{\alpha\beta} / 2)^2\}^{1/2} \quad (15)$$

and then minimize in Q to obtain

$$Q_0 = \sqrt{(2E_{JT}/K)} \left[1 - (E_{\alpha\beta}/4E_{JT})^2\right]^{1/2} \tag{16}$$

which is the reorientational off-center radius for $4E_{JT} \geq E_{\alpha\beta}$. Hereafter

$$E_{JT} = b^2 / 2M\omega^2 \tag{17}$$

stands for Jahn-Teller's energy of the off-centering process.

Inserting $Q = Q_0$ into (15) we get the lower-branch vibronic Hamiltonian for free rotation upon the off-centered sphere with $\sum_i Q_i^2 = Q_0^2$:

$$H_{vib0}(d=0) = -(\hbar^2/2I)\Delta_{\theta\varphi} - E_{JT}\left[1 + (E_{\alpha\beta}/4E_{JT})^2\right]$$

where we have introduced spherical coordinates

$$Q_x = Q_0 \cos\varphi \sin\theta$$

$$Q_y = Q_0 \sin\varphi \sin\theta$$

$$Q_z = Q_0 \cos\theta$$

$$\Delta_{\varphi\theta} = (1/\sin\theta)\{(\partial/\partial\theta)[\sin\theta(\partial/\partial\theta)] + (\partial/\partial\varphi)[(1/\sin\theta)(\partial/\partial\varphi)]$$

as well as the inertial moment of the off-centered entity

$$I = MQ_0^2$$

The quantized energy of free rotation is [6]

$$E_n = \hbar^2 n(n+1)/2MQ_0^2 = \hbar^2 n(n+1)/2I \tag{18}$$

Rehabilitating the hindering terms, the d-dependent correction to eq. (15) being small equation (14) converts to

$$E_{U/L}(Q_0) = \pm\left[(d_c - d_b)\sum_i Q_i^4 + d_b Q_0^4\right](M\omega^2/b) + E_{JT}\left[(1\pm2) - (E_{\alpha\beta}/4E_{JT})^2\right] \tag{19}$$

Equation (19) defines the potential energy surface controlling the hindered rotation upon a reorientational sphere. Indeed from (13) we get

$$H_{vib0}(d\neq 0) = -(\hbar^2/2M)\sum_i(\partial^2/\partial Q_i^2) \pm (M\omega^2/b)[(d_c - d_b)\sum_i Q_i^4 + d_b Q_0^4\}] +$$

$$E_{JT}\left[(1\pm2) - (E_{\alpha\beta}/4E_{JT})^2\right] \tag{20}$$

or equivalently

$$H_{vib0}(d \neq 0) = -(\hbar^2/2I)\Delta(\theta,\varphi) \pm (M\omega^2/b) Q_0^4 (d_c - d_b)[(\cos\varphi \sin\theta)^4 + (\sin\varphi \sin\theta)^4 +$$

$$(\cos\theta)^4] + d_b + E_{JT}[(1\pm 2) - (E_{\alpha\beta}/4E_{JT})^2] \tag{21}$$

with $\sum_i Q_i^2 = Q_0^2$. We see that the rotation upon the off-centered sphere is hindered by terms fourth power in $Q_i$. The latter terms are obliged to the asymmetry between $d_c$ and $d_b$. Apart from the nonessential constant terms, the (20)-based Schrödinger equation splits into three equivalent but interdependent eigenvalue equations, one for each $Q_i$.

Reorientational sites are the minima of $E_L(Q_0)$ with respect to $Q_i$, while the maxima occur as saddle points in between. Both extrema appear by virtue of the asymmetry between $d_c$ and $d_b$. To sort them out, eq. (19) is to be minimized under the condition $Q_0 = \sqrt{(\sum_i Q_i^2)}$. We introduce Laplace's multiplier $\mu_i$ and differentiate equation (19) in $Q_i$ at constant $Q_0$ to get:

$$\partial E_L / \partial Q_i - \mu_i (Q_i / Q_0) \equiv -\{4(M\omega^2/b)(d_c - d_b) Q_i^3 + \mu_i (Q_i / Q_0)\} = 0$$

There is a root $Q_i = 0$ and two other ones given by

$$Q_i = \pm [\mu_i / 4 Q_0 (M\omega^2/b)(d_b - d_c)]^{1/2}$$

The Laplace multipliers $\mu_i$ are determined from $Q_0^2 = \sum_i Q_i^2$ giving

$$\sum_i \mu_i = 4 Q_0^3 (M\omega^2/b)(d_b - d_c)$$

where i=x,y,z. We solve for $\mu_i$ in three distinct cases relevant to the assumed crystalline geometry.

In <100> symmetry, e.g. $\mu_x \neq 0$, $\mu_y = \mu_z = 0$, a configurational $Q_x$-axis should host a site: $Q_x = Q_0$, $Q_y = Q_z = 0$. We have

$$\mu_x = 4(M\omega^2/b)(d_b - d_c) Q_0^3, \quad \mu_y = \mu_z = 0$$

Alternatively $\mu_i$ may be determined in <111> symmetry setting $\mu_x = \mu_y = \mu_z$, leading to sites at $Q_i = Q_0 / \sqrt{3}$ (i=x,y,z):

$$\mu_i = (4/3)(M\omega^2/b)(d_b - d_c) Q_0^3, \quad (i=x,y,z).$$

A third possibility is to determine $\mu_i$ in <110> symmetry, e.g. by $\mu_x = \mu_y$, $\mu_z = 0$, leading to sites at $Q_x = Q_y = Q_0 / \sqrt{2}$, $Q_z = 0$:

$$\mu_x = \mu_y = 2(M\omega^2/b)(d_b - d_c) Q_0^3, \quad \mu_z = 0$$

By taking a second derivative in $Q_i$, we get minima and maxima according to whether positive or negative:

$$\partial^2 E_L / \partial Q_i^2 \equiv 12 (M\omega^2 / b)(d_b - d_c) Q_i^2 = 3(\mu_i / Q_0)$$

$$= 12 (M\omega^2 / b)(d_b - d_c) Q_0^2 > 0, \text{ six} <100> \text{ sites, or}$$

$$= 4 (M\omega^2 / b)(d_b - d_c) Q_0^2 > 0, \text{ eight} <111> \text{ sites, or}$$

$$= 6 (M\omega^2 / b)(d_b - d_c) Q_0^2 > 0, \text{ twelve} <110> \text{ sites,}$$

all being minima for $d_b - d_c > 0$ (though maxima for $d_b - d_c < 0$).

From eq.(20) the spatial curvature of the angle-dependent part

$$\pm (M\omega^2 / b)(d_c - d_b)[(Q_0 \cos\varphi \sin\theta)^4 + (Q_0 \sin\varphi \sin\theta)^4 + (Q_0 \cos\theta)^4]$$

of the vibronic potentials along $Q_x = Q_0 \cos\varphi \sin\theta$, $Q_y = Q_0 \sin\varphi \sin\theta$, $Q_z = Q_0 \cos\theta$ is, respectively,

$$\partial^2 E_{U/L} / \partial Q_x^2 = \pm 12(I\omega^2)[(d_c - d_b) / b](\cos\varphi \sin\theta)^2$$

$$\partial^2 E_{U/L} / \partial Q_x^2 = \pm 12(I\omega^2)[(d_c - d_b) / b](\sin\varphi \sin\theta)^2$$

$$\partial^2 E_{U/L} / \partial Q_x^2 = \pm 12(I\omega^2)[(d_c - d_b) / b](\cos\theta)^2$$

giving rise to the following effective force constants defined by

$$K_{eff} = (1/3)[(\partial^2 E_{U/L} / \partial Q_x^2) + (\partial^2 E_{U/L} / \partial Q_y^2) + (\partial^2 E_{U/L} / \partial Q_z^2)]:$$

$$K_{min} = + 4 (I\omega^2)[(d_b - d_c) / b] = + 4 K Q_0^2 [(d_b - d_c) / b] \qquad (22)$$

at the well bottoms and

$$K_{max} = - 4 (I\omega^2)[(d_b - d_c) / b] = - 4 K Q_0^2 [(d_b - d_c) / b] \qquad (23)$$

at the interwell tops. Introducing

$$K = M\omega^2 \equiv M\omega_{bare}^2$$

and

$$K_{min} \equiv K_{ren} = M\omega_{renII}^2,$$

we define a renormalized rotational frequency

$$\omega_{renII} = \omega_{bare} [4 (d_b - d_c)/b]^{1/2} Q_0 = \omega_{renI} [8E_{JT}(d_b - d_c)/bK]^{1/2} \quad (24)$$

where

$$\omega_{renI} = \omega_{bare} [1 - (E_{\alpha\beta}/4E_{JT})^2]^{1/2} \quad (25)$$

is the renormalized off-centering frequency [3] with $\omega_{bare} \equiv \omega$ standing for the bare vibrational frequency. From equation (16), at large $E_{\alpha\beta} \approx 2 b^2 / M\omega_{bare}^2$, $Q_0$ is nearly vanishing and so is $\omega_{renII}$. At small $E_{\alpha\beta}$, $Q_0 \approx b/K = b/(M\omega_{bare}2)$, so that $\omega_{renII}$ varies as $\omega_{bare}^{-1}$. On introducing $\omega_{renII}$, the angle-dependent part of the vibronic potential becomes

$$\pm \tfrac{1}{4} (M\omega_{ren}^2) [(Q_0 \cos\varphi \sin\theta)^4 + (Q_0 \sin\varphi \sin\theta)^4 + (Q_0 \cos\theta)^4]$$

### 3.3. Mixing constants

As the particular ion is driven off-site, its electrostatic potential in the point-ion field modulated by the displacements $Q_k$ is

$$U(r_0, Q_l) = \alpha_M e / [(x_0 + Q_x)^2 + (y_0 + Q_y)^2 + (z_0 + Q_z)^2]^{1/2}. \quad (26)$$

$\alpha_M$, $V_M = \alpha_M e^2 / r_0$, and $r_0 = (x_0^2 + y_0^2 + g_0^2)^{1/2}$ are Madelung's constant, potential and 'cavity radius' at the impurity site. $U(r_0, Q_l)$ generates an electric field $F(r_0, Q_l) = -\text{grad}_Q U(r_0, Q_l)$. This field couples to the electric dipole e**r** which mixes the electronic states involved. The coupling energy $V = - e\, \mathbf{r}.\mathbf{F}$ is

$$V(r_0, Q_l) = - (\alpha_M e) [ex(x_0 + Q_x) + ex(x_0 + Q_x) + ex(x_0 + Q_x)]/$$

$$[(x_0 + Q_x)^2 + (y_0 + Q_y)^2 + (z_0 + Q_z)^2]^{3/2} \quad (27)$$

To derive electron-phonon coupling operators, we differentiate $V(r_0, Q_l)$. Namely, setting

$$x/r = x_0/r_0 = \cos\varphi \sin\theta$$

$$y/r = y_0/r_0 = \sin\varphi \sin\theta$$

$$z/r = z_0/r_0 = \cos\theta$$

and dropping for simplicity the $\alpha_M e^2$ factor we get:

$$V(r_0, Q_l)|_{Q=0} = r/r_0^2$$

$\partial V(r_0, Q_l) / \partial Q_x |_{Q=0} = -2x / r_0^3$

$\partial V(r_0, Q_l) / \partial Q_x \partial Q_y |_{Q=0} = 9 x y_0 / r_0^5$

$\partial^2 V(r_0, Q_l) / \partial Q_x^2 |_{Q=0} = 3 (r r_0 + 3x x_0) / r_0^5$

$\partial^3 V(r_0, Q_l) / \partial Q_x \partial Q_y \partial Q_z |_{Q=0} = 74 x y_0 z_0 / r_0^7$

$\partial^3 V(r_0, Q_l) / \partial Q_x^2 \partial Q_z |_{Q=0} = -6 x_0 (r r_0 + 11 x x_0) / r_0^7$

$\partial^3 V(r_0, Q_l) / \partial Q_x^3 |_{Q=0} = 6 x (3 r_0^2 - 11 x_0^2) / r_0^7$

For estimating the coupling constants

$v_{ij...k,xyz} = <t_{1u,xyz} | v_{ij...k}(r) | a_{1g}> \sim \int dr\, d\theta\, d\varphi\, r_{xyz}\, v_{ij...k}(r) \exp[-(\alpha_{t1u} + \alpha_{a1g})r] r^2 \sin\theta$

$= \int_0^\infty dr \int_0^{2\pi} d\varphi \int_0^\pi d\theta\, r_{xyz}\, v_{ij...k}(r)\, r^2 \sin\theta \exp[-(\alpha_{t1u} + \alpha_{a1g})r]$

we use hydrogen-like atomic wavefunctions [7]

$|a_{1g}> = \pi^{-1/2} (Z/a_0)^{3/2} \exp(-Zr/a_0) \equiv N_{a1g} \exp(-\alpha_{a1g} r)$

$|t_{1u,x}> = (32\pi)^{-1/2} (Z/a_0)^{3/2} (Zr/a_0) \exp(-Zr/2a_0) \cos\varphi \sin\theta \equiv N_{t1u}\, x \exp(-\alpha_{t1u} r)$

$|t_{1u,x}> = (32\pi)^{-1/2} (Z/a_0)^{3/2} (Zr/a_0) \exp(-Zr/2a_0) \sin\varphi \sin\theta \equiv N_{t1u}\, y \exp(-\alpha_{t1u} r)$

$|t_{1u,x}> = (32\pi)^{-1/2} (Z/a_0)^{3/2} (Zr/a_0) \exp(-Zr/2a_0) \cos\theta \equiv N_{t1u}\, z \exp(-\alpha_{t1u} r)$

$N_{a1g} = \pi^{-1/2} (Z/a_0)^{3/2}$

$N_{t1u} = (32\pi)^{-1/2} (Z/a_0)^{5/2}$

$\alpha_{a1g} = Z/a_0$

$\alpha_{t1u} = Z/2a_0$

$a_0 = h^2 / 4\pi^2 \mu e^2$ being Bohr's radius. We calculate (see Appendix I)

$d_{xxx,x} = -10.39 [\alpha_M e^2 / (ar_0)^5] a^4$ for $d_c$

$d_{xxz,z} = +21.45 [\alpha_M e^2 / (ar_0)^5] a^4$ for $d_b$ and

$b_{x,x} = 2.23 [\alpha_M e^2 / (ar_0)^3] a^2$ for b

with $a = \alpha_{a1g} + \alpha_{t1u}$ and therefore $d_b > d_c$. At the same time, the remaining constants are all vanishing as the model requires: $d_{xxx,y}$, $d_{xxx,z}$, $d_{xxz,x}$, $d_{xxz,y}$, $d_{xyz,x}$, $d_{xyz,y}$, $d_{xyz,z}$, $c_{xx,x}$, $c_{xx,y}$, $c_{xx,z}$, $c_{xy,x}$, $c_{xy,y}$, $c_{xy,z}$, $b_{x,y}$, $b_{x,z}$.

## 4. Conclusion

We confirmed analytically that while off-center displacements of a substitutional Li impurity in an alkali halide crystal arose from the first-order electron-mode coupling, rotational barriers resulted from the third-order coupling to the $T_{1u}$ vibrational mode of the three halogen pairs centered at the respective cation site. Applying to a specific model, we calculated coupling constants and thereby potential-energy profiles with periodic barriers between reorientational wells. We also found that the hindered reorientation of the off-centered Li impurity across the barriers required renormalization of the vibrational frequency additional to the one which controlled the off-center displacement.

## Appendix I

### Calculation of the mixing constants

Omitting for simplicity the normalization factors $N_{a1g} N_{t1u} = (1 / 4 \pi \sqrt{2})(Z / a_0)^4$, we get

$$d_{xxx,x} \sim (\alpha_M e^2 / r_0^7) \int_0^\infty dr \int_0^{2\pi} d\varphi \int_0^\pi d\theta\, 6x^2(3r_0^2 - 11x_0^2)(r^2 \sin\theta)\exp[-(\alpha_{t1u} + \alpha_{a1g})r]$$

$$= (\alpha_M e^2 / r_0^7) \int_0^\infty dr \int_0^{2\pi} d\varphi \int_0^\pi d\theta\, 6r^2 (\cos\varphi \sin\theta)^2 (r^2 \sin\theta) r_0^2 \times$$

$[3 - 11(\cos\varphi \sin\theta)^2 ]\exp[-(\alpha_{t1u} + \alpha_{a1g})r ]$

$d_{xxx,y} \sim (\alpha_M e^2 / r_0^7 )\, _0\!\int^\infty dr\, _0\!\int^{2\pi} d\varphi\, _0\!\int^\pi d\theta\, 6xy (3r_0^2 - 11x_0^2 )(r^2 \sin\theta)\exp[-(\alpha_{t1u} + \alpha_{a1g})r ]$

$= (\alpha_M e^2 / r_0^7 )\, _0\!\int^\infty dr\, _0\!\int^{2\pi} d\varphi\, _0\!\int^\pi d\theta\, 6r^2 (\cos\varphi \sin\varphi)(\sin\theta)^2 (r^2 \sin\theta)\, r_0^2 \times$

$[3 - 11(\cos\varphi \sin\theta)^2 ]\exp[-(\alpha_{t1u} + \alpha_{a1g})r ] = 0$

$d_{xxx,z} \sim (\alpha_M e^2 / r_0^7 )\, _0\!\int^\infty dr\, _0\!\int^{2\pi} d\varphi\, _0\!\int^\pi d\theta\, 6xz (3r_0^2 - 11x_0^2 )(r^2 \sin\theta)\exp[-(\alpha_{t1u} + \alpha_{a1g})r ]$

$= (\alpha_M e^2 / r_0^7 )\, _0\!\int^\infty dr\, _0\!\int^{2\pi} d\varphi\, _0\!\int^\pi d\theta\, 6r^2 (\cos\varphi \sin\theta)(\cos\theta)(r^2 \sin\theta)\, r_0^2 \times$

$[3 - 11(\cos\varphi \sin\theta)^2 ]\exp[-(\alpha_{t1u} + \alpha_{a1g})r ] = 0$

$d_{xxz,x} \sim -(\alpha_M e^2 / r_0^7 )\, _0\!\int^\infty dr\, _0\!\int^{2\pi} d\varphi\, _0\!\int^\pi d\theta\, 6xz_0 (rr_0 + 11zx_0 )(r^2\sin\theta)\exp[-(\alpha_{t1u} + \alpha_{a1g})r ]$

$= -(\alpha_M e^2 / r_0^7 )\, _0\!\int^\infty dr\, _0\!\int^{2\pi} d\varphi\, _0\!\int^\pi d\theta\, 6rr_0 (\cos\varphi \sin\theta)(\cos\theta)(r^2 \sin\theta)\, rr_0 \times$

$[1 + 11(\cos\varphi \sin\theta)^2 ]\exp[-(\alpha_{t1u} + \alpha_{a1g})r ] = 0$

$d_{xxz,y} \sim -(\alpha_M e^2 / r_0^7 )\, _0\!\int^\infty dr\, _0\!\int^{2\pi} d\varphi\, _0\!\int^\pi d\theta\, 6yz_0 (rr_0 + 11zx_0 )(r^2\sin\theta)\exp[-(\alpha_{t1u} + \alpha_{a1g})r ]$

$= -(\alpha_M e^2 / r_0^7 )\, _0\!\int^\infty dr\, _0\!\int^{2\pi} d\varphi\, _0\!\int^\pi d\theta\, 6rr_0 (\sin\varphi \sin\theta)(\cos\theta)(r^2 \sin\theta)\, rr_0 \times$

$[1 + 11(\cos\varphi \sin\theta)^2 ]\exp[-(\alpha_{t1u} + \alpha_{a1g})r ] = 0$

$d_{xxz,z} \sim -(\alpha_M e^2 / r_0^7 )\, _0\!\int^\infty dr\, _0\!\int^{2\pi} d\varphi\, _0\!\int^\pi d\theta\, 6zz_0 (rr_0 + 11zx_0 )(r^2\sin\theta)\exp[-(\alpha_{t1u} + \alpha_{a1g})r ]$

$= -(\alpha_M e^2 / r_0^7 )\, _0\!\int^\infty dr\, _0\!\int^{2\pi} d\varphi\, _0\!\int^\pi d\theta\, 6rr_0 (\cos\theta)^2 (r^2 \sin\theta)\, rr_0 \times$

$[1 + 11(\cos\varphi \sin\theta)^2 ]\exp[-(\alpha_{t1u} + \alpha_{a1g})r ]$

$d_{xyz,x} \sim (\alpha_M e^2 / r_0^7 )\, _0\!\int^\infty dr\, _0\!\int^{2\pi} d\varphi\, _0\!\int^\pi d\theta\, 74x^2 y_0 z_0 (r^2 \sin\theta) \exp[-(\alpha_{t1u} + \alpha_{a1g})r ]$

$= (\alpha_M e^2 / r_0^7 )\, _0\!\int^\infty dr\, _0\!\int^{2\pi} d\varphi\, _0\!\int^\pi d\theta\, 74(rr_0 )^2 (\cos\varphi \sin\theta)^2 (\sin\varphi \sin\theta)(\cos\theta)(r^2 \sin\theta) \times$

$\exp[-(\alpha_{t1u} + \alpha_{a1g})r ] = 0$

$d_{xyz,y} \sim (\alpha_M e^2 / r_0^7 )\, _0\!\int^\infty dr\, _0\!\int^{2\pi} d\varphi\, _0\!\int^\pi d\theta\, 74x\, yy_0 z_0 (r^2 \sin\theta) \exp[-(\alpha_{t1u} + \alpha_{a1g})r ]$

$= (\alpha_M e^2 / r_0^7 )\, _0\!\int^\infty dr\, _0\!\int^{2\pi} d\varphi\, _0\!\int^\pi d\theta\, 74(rr_0 )^2 (\sin\varphi \sin\theta)^2 (\cos\varphi \sin\theta)(\cos\theta)(r^2 \sin\theta) \times$

$\exp[-(\alpha_{t1u} + \alpha_{a1g})r ] = 0$

$$d_{xyz,z} \sim (\alpha_M e^2 / r_0^7)\, {}_0\!\int^\infty dr\, {}_0\!\int^{2\pi} d\varphi\, {}_0\!\int^\pi d\theta\, 74zx\, y_0 z_0\, (r^2 \sin\theta) \exp[-(\alpha_{t1u} + \alpha_{a1g})r]$$

$$= (\alpha_M e^2 / r_0^7)\, {}_0\!\int^\infty dr\, {}_0\!\int^{2\pi} d\varphi\, {}_0\!\int^\pi d\theta\, 74(rr_0)^2 (\cos\varphi \sin\theta)(\sin\varphi \sin\theta)(\cos\theta)^2 (r^2 \sin\theta) \times$$

$$\exp[-(\alpha_{t1u} + \alpha_{a1g})r] = 0$$

$$c_{xx,x} \sim (\alpha_M e^2 / r_0^5)\, {}_0\!\int^\infty dr\, {}_0\!\int^{2\pi} d\varphi\, {}_0\!\int^\pi d\theta\, 3x (rr_0 + 3xx_0)(r^2 \sin\theta) \exp[-(\alpha_{t1u} + \alpha_{a1g})r]$$

$$= (\alpha_M e^2 / r_0^5)\, {}_0\!\int^\infty dr\, {}_0\!\int^{2\pi} d\varphi\, {}_0\!\int^\pi d\theta\, 3r^2 r_0 (\cos\varphi \sin\theta)(r^2 \sin\theta) \times$$

$$[1 + 3(\cos\varphi \sin\theta)^2] \exp[-(\alpha_{t1u} + \alpha_{a1g})r] = 0$$

$$c_{xx,y} \sim (\alpha_M e^2 / r_0^5)\, {}_0\!\int^\infty dr\, {}_0\!\int^{2\pi} d\varphi\, {}_0\!\int^\pi d\theta\, 3y (rr_0 + 3xx_0)(r^2 \sin\theta) \exp[-(\alpha_{t1u} + \alpha_{a1g})r]$$

$$= (\alpha_M e^2 / r_0^5)\, {}_0\!\int^\infty dr\, {}_0\!\int^{2\pi} d\varphi\, {}_0\!\int^\pi d\theta\, 3r^2 r_0 (\sin\varphi \sin\theta)(r^2 \sin\theta) \times$$

$$[1 + 3(\cos\varphi \sin\theta)^2] \exp[-(\alpha_{t1u} + \alpha_{a1g})r] = 0$$

$$c_{xx,z} \sim (\alpha_M e^2 / r_0^5)\, {}_0\!\int^\infty dr\, {}_0\!\int^{2\pi} d\varphi\, {}_0\!\int^\pi d\theta\, 3z (rr_0 + 3xx_0)(r^2 \sin\theta) \exp[-(\alpha_{t1u} + \alpha_{a1g})r]$$

$$= (\alpha_M e^2 / r_0^5)\, {}_0\!\int^\infty dr\, {}_0\!\int^{2\pi} d\varphi\, {}_0\!\int^\pi d\theta\, 3r^2 r_0 (\cos\theta)(r^2 \sin\theta) \times$$

$$[1 + 3(\cos\varphi \sin\theta)^2] \exp[-(\alpha_{t1u} + \alpha_{a1g})r] = 0$$

$$c_{xy,x} \sim (\alpha_M e^2 / r_0^5)\, {}_0\!\int^\infty dr\, {}_0\!\int^{2\pi} d\varphi\, {}_0\!\int^\pi d\theta\, 9x^2 y_0 (r^2 \sin\theta) \exp[-(\alpha_{t1u} + \alpha_{a1g})r]$$

$$= (\alpha_M e^2 / r_0^5)\, {}_0\!\int^\infty dr\, {}_0\!\int^{2\pi} d\varphi\, {}_0\!\int^\pi d\theta\, 9r^2 r_0 (\cos\varphi \sin\theta)(r^2 \sin\theta) \times$$

$$(\sin\varphi \sin\theta)^2 \exp[-(\alpha_{t1u} + \alpha_{a1g})r] = 0$$

$$c_{xy,y} \sim (\alpha_M e^2 / r_0^5)\, {}_0\!\int^\infty dr\, {}_0\!\int^{2\pi} d\varphi\, {}_0\!\int^\pi d\theta\, 9xyy_0 (r^2 \sin\theta) \exp[-(\alpha_{t1u} + \alpha_{a1g})r]$$

$$= (\alpha_M e^2 / r_0^5)\, {}_0\!\int^\infty dr\, {}_0\!\int^{2\pi} d\varphi\, {}_0\!\int^\pi d\theta\, 9r^2 r_0 (\cos\varphi \sin\theta)(r^2 \sin\theta) \times$$

$$(\sin\varphi \sin\theta)^2 \exp[-(\alpha_{t1u} + \alpha_{a1g})r] = 0$$

$$c_{xy,z} \sim (\alpha_M e^2 / r_0^5)\, {}_0\!\int^\infty dr\, {}_0\!\int^{2\pi} d\varphi\, {}_0\!\int^\pi d\theta\, 9xzy_0 (r^2 \sin\theta) \exp[-(\alpha_{t1u} + \alpha_{a1g})r]$$

$$= (\alpha_M e^2 / r_0^5)\, {}_0\!\int^\infty dr\, {}_0\!\int^{2\pi} d\varphi\, {}_0\!\int^\pi d\theta\, 9r^2 r_0 (\cos\varphi \sin\theta)(\sin\varphi \sin\theta)\cos\theta\, (r^2 \sin\theta) \times$$

$$\exp[-(\alpha_{t1u} + \alpha_{a1g})r] = 0$$

$$b_{x,x} \sim -(\alpha_M e^2 / r_0^3)\, {}_0\!\int^\infty dr\, {}_0\!\int^{2\pi} d\varphi\, {}_0\!\int^\pi d\theta\, 2x^2 (r^2 \sin\theta) \exp[-(\alpha_{t1u} + \alpha_{a1g})r]$$

$$= - (\alpha_M e^2 / r_0^3) \,_0\!\int^\infty dr \,_0\!\int^{2\pi} d\varphi \,_0\!\int^\pi d\theta \, 2r^2 (\cos\varphi \sin\theta)^2 (r^2 \sin\theta) \exp[-(\alpha_{t1u} + \alpha_{a1g})r]$$

$$b_{x,y} \sim - (\alpha_M e^2 / r_0^3) \,_0\!\int^\infty dr \,_0\!\int^{2\pi} d\varphi \,_0\!\int^\pi d\theta \, 2xy (r^2 \sin\theta) \exp[-(\alpha_{t1u} + \alpha_{a1g})r]$$

$$= - (\alpha_M e^2 / r_0^3) \,_0\!\int^\infty dr \,_0\!\int^{2\pi} d\varphi \,_0\!\int^\pi d\theta \, 2r^2 (\cos\varphi \sin\theta)(\sin\varphi \sin\theta)(r^2 \sin\theta) \times$$

$$\exp[-(\alpha_{t1u} + \alpha_{a1g})r] = 0$$

$$b_{x,z} \sim - (\alpha_M e^2 / r_0^3) \,_0\!\int^\infty dr \,_0\!\int^{2\pi} d\varphi \,_0\!\int^\pi d\theta \, 2xz (r^2 \sin\theta) \exp[-(\alpha_{t1u} + \alpha_{a1g})r]$$

$$= - (\alpha_M e^2 / r_0^3) \,_0\!\int^\infty dr \,_0\!\int^{2\pi} d\varphi \,_0\!\int^\pi d\theta \, 2r^2 (\cos\varphi \sin\theta)(\cos\theta)(r^2 \sin\theta) \times$$

$$\exp[-(\alpha_{t1u} + \alpha_{a1g})r] = 0$$

the vanishing ones following integration in $\varphi$ or $\theta$. The integration in r is done using

$$_0\!\int^\infty r^n \exp(-ar)\, dr = (-1)^{n+1} (n! / a^{n+1}).$$

We get setting $a = \alpha_{t1u} + \alpha_{a1g} = (3/2)(Z/a_0)$

$$b_{x,x} \sim - (\alpha_M e^2 / r_0^3) \,_0\!\int^\infty dr \,_0\!\int^{2\pi} d\varphi \,_0\!\int^\pi d\theta \, 2r^2 (\cos\varphi \sin\theta)^2 (r^2 \sin\theta) \exp[-(\alpha_{t1u} + \alpha_{a1g})r]$$

$$\times (1/4\pi\sqrt{2})(Z/a_0) = 2.23481 \,[\alpha_M e^2 / (ar_0)^3]\, a^2$$

$$d_{xxx,x} \sim (\alpha_M e^2 / r_0^7) \,_0\!\int^\infty dr \,_0\!\int^{2\pi} d\varphi \,_0\!\int^\pi d\theta \, 6r^2 (\cos\varphi \sin\theta)^2 (r^2 \sin\theta)\, r_0^2 \times$$

$$[3 - 11(\cos\varphi \sin\theta)^2] \exp[-(\alpha_{t1u} + \alpha_{a1g})r] (1/4\pi\sqrt{2})(Z/a_0)^4$$

$$= -10.39185 \,[\alpha_M e^2 / (ar_0)^5]\, a^4$$

$$d_{xxz,z} \sim - (\alpha_M e^2 / r_0^7) \,_0\!\int^\infty dr \,_0\!\int^{2\pi} d\varphi \,_0\!\int^\pi d\theta \, 6rr_0 (\cos\theta)^2 (r^2 \sin\theta)\, rr_0 \times$$

$$[1 + 11(\cos\varphi \sin\theta)^2] \exp[-(\alpha_{t1u} + \alpha_{a1g})r] (1/4\pi\sqrt{2})(Z/a_0)^4$$

$$= 21.45414 \,[\alpha_M e^2 / (ar_0)^5]\, a^4$$

## Appendix II

Tables of numerical parameters

Table I

Calculated coupling parameters

| Host | Cavity radius $r_0$ (Å)[a,c] | Madel. energy $V_M$ (eV)[b,c] | $k_o$[c] | $k_p$[c] | $k_s$[c] | $b$[d] (eV/Å) | $d_c$[d] (eV/Å³) | $d_b$[d] (eV/Å³) |
|---|---|---|---|---|---|---|---|---|
| LiF  | 2.014 | 12.37 | 1.96 | 2.50 | 9.01  | 7.6715 | -1.4658 | 3.0261 |
| NaF  | 2.317 | 10.77 | 1.74 | 2.65 | 5.05  | 4.4801 | -0.6468 | 1.3352 |
| KF   | 2.674 | 9.33  | 1.85 | 2.80 | 5.46  | 3.0982 | -0.3358 | 0.6933 |
| RbF  | 2.815 | 8.81  | 1.96 | 2.81 | 6.48  | 2.7967 | -0.2735 | 0.5647 |
| LiCl | 2.570 | 9.68  | 2.78 | 3.62 | 11.95 | 5.2291 | -0.6136 | 1.2667 |
| NaCl | 2.820 | 8.86  | 2.34 | 3.88 | 5.90  | 3.3460 | -0.3261 | 0.6732 |
| KCl  | 3.147 | 7.94  | 2.19 | 4.00 | 4.84  | 2.2534 | -0.1763 | 0.3641 |
| RbCl | 3.291 | 7.64  | 2.19 | 3.95 | 4.92  | 1.9827 | -0.1419 | 0.2929 |
| LiBr | 2.751 | 9.02  | 3.17 | 4.17 | 13.25 | 4.8491 | -0.4966 | 1.0252 |
| NaBr | 2.989 | 8.37  | 2.59 | 4.41 | 6.28  | 3.1142 | -0.2701 | 0.5577 |
| KBr  | 3.298 | 7.58  | 2.34 | 4.48 | 4.90  | 2.0929 | -0.1491 | 0.3079 |
| RbBr | 3.445 | 7.26  | 2.34 | 4.51 | 4.86  | 1.8372 | -0.1200 | 0.2477 |
| LiI  | 3.000 | 8.19  | 3.80 | 4.90 | 16.85 | 4.4381 | -0.3822 | 0.7890 |
| NaI  | 3.237 | 7.73  | 2.93 | 4.90 | 7.28  | 2.7742 | -0.2052 | 0.4236 |
| KI   | 3.533 | 7.06  | 2.62 | 5.39 | 5.10  | 1.9019 | -0.1181 | 0.2438 |
| RbI  | 3.671 | 6.79  | 2.59 | 5.48 | 4.91  | 1.6748 | -0.0963 | 0.1989 |

[a]The effective cavity radius is usually taken to be $r_0 \sim r_{ac}$, the nearest-neighbor cation-anion separation in an fcc lattice. [b]The Madelung energy is $V_M = a_M e^2 / r_0$. [c]Data from W. Beall Fowler, Physics of Color Centers (Academic, New York, 1968). [d]We use an enlarged Bohr-orbit radius $a_0 = (h^2/2\mu)(k/2\pi^2 e^2) = 1.34 \times 10^{-2}\, k$ (Å) proportional to the dielectric constant k, optical $k_o$ or polaronic $k_p = [k_o^{-1} - k_s^{-1}]^{-1}$ where $k_s$ is the static constant. We calculate $Z / a_0 = (2/3) a k$ by equating $b = 2.2348[V_M / r_0 (ar_0)]$ to KCl data $G = 2.2534$ eV/Å from Table II by Baldacchini et al., Nuovo Cim. **13D**, 1399 (1991). Given $r_0$, this equation yields $a$. Z is the effective charge +0.34e on the Li$^+$ ion from an extended Hückel analysis of a LiCl$_6$ cluster. The computed $Z / a_0 = 1.1608$ CGSE / Å is assumed to hold good for all alkali halide hosts. We also checked the feasibility of another cavity radius $r_0 = r_{Li+}$ which, however, led to unrealistic Z.

Table II

Calculated dynamic prameters

| Host | Mode mass $M$[e] (at.u.) | Attempt frequency $\omega_R$[f,c] ($10^{13}$ s$^{-1}$) | Force constant $K = M\omega_R^2$ (eV/Å²) | JT energy $E_{JT}$ (eV) | Electron energy gap $E_{\alpha\beta}$ (eV) | $E_{\alpha\beta}/4E_{JT}$ |
|---|---|---|---|---|---|---|

| Host | | | | | | |
|---|---|---|---|---|---|---|
| LiF | 28.4976 | 5.78 | 9.9235 | 2.9653 | 1.602[g] | 0.1351 |
| NaF | " | 4.64 | 6.3951 | 1.5693 | " | 0.2552 |
| KF | " | 3.58 | 3.8069 | 1.2607 | " | 0.3177 |
| RbF | " | 2.94 | 2.5675 | 1.5232 | " | 0.2629 |
| LiCl | 53.1795 | 3.61 | 7.2237 | 1.8926 | 6.779[g] | 0.8955 |
| " | " | " | " | " | 1.406[i] | 0.1857 |
| NaCl | " | 3.20 | 5.6761 | 0.9862 | 1.406[i] | 0.3564 |
| KCl | " | 2.81 | 4.3769 | 0.5801 | " | 0.6059 |
| RbCl | " | 2.39 | 3.1662 | 0.6208 | " | 0.5662 |
| LiBr | 119.8560 | - | - | - | 1.237[h] | - |
| NaBr | " | 2.55 | 8.1235 | 0.5969 | " | 0.5181 |
| KBr | " | 2.17 | 5.8828 | 0.3723 | " | 0.8306 |
| RbBr | " | 1.82 | 4.1381 | 0.4078 | " | 0.7583 |
| LiI | 190.3568 | - | - | - | 1.379[h] | - |
| NaI | " | 2.20 | 9.6032 | 0.4007 | " | 0.8604 |
| KI | " | 2.04 | 8.2571 | 0.2190 | " | 1.5742 |
| RbI | " | 1.54 | 4.7056 | 0.2980 | " | 1.1569 |

[e] $M=1.5 M_x$ where $M_x$ is the halogen mass (cf. Baldacchini et al. 1991). [f] $\omega_R$ are the Restrahlen frequencies. [g]Calculated as $E_{3a1g} - E_{4t1u}$ of a LiHal$_6$ cluster. [h]Calculated as $E_{2a1g} - E_{4t1u}$ of a LiHal$_6$ cluster. Average of $E_{\alpha\beta}$ data under [g] and [h] for Hal=F,Br,I.

Table III

Calculated off-center displacement parameters

| Host | Off-center radius $Q_0$ (Å) | I ren. frequency $\omega_{renI}$ ($10^{13}$ s$^{-1}$) | Off-on barrier $E_{BI}$ (eV) | Lattice relaxation energy $E_{RI}$[j] (eV) | Crossover energy $E_{CI}$[k] (eV) | Optical energy $E_{OI}$[l] (eV) |
|---|---|---|---|---|---|---|
| LiF | 0.7660 | 5.7270 | 2.2182 | 11.6453 | 3.0192 | 11.8612 |
| NaF | 0.6774 | 4.4864 | 0.8705 | 5.8690 | 1.6715 | 6.2772 |
| KF | 0.7717 | 3.3945 | 0.5869 | 4.5342 | 1.3879 | 5.0428 |
| RbF | 1.0510 | 2.8366 | 0.8276 | 5.6721 | 1.6286 | 6.0928 |
| LiCl | 0.3222 | 1.6067 | 0.0207 | 1.4998 | 3.4102 | 7.5704 |
| " | 0.7113 | 3.5472 | 1.2550 | 7.3096 | 1.9580 | " |
| NaCl | 0.5508 | 2.9899 | 0.4085 | 3.4440 | 1.1115 | 3.9448 |
| KCl | 0.4096 | 2.2355 | 0.0901 | 1.4686 | 0.7931 | 2.3204 |
| RbCl | 0.5162 | 1.9700 | 0.1168 | 1.6873 | 0.8198 | 2.4832 |
| LiBr | -      | -      | -      | -      | -      | -      |
| NaBr | 0.3279 | 2.1811 | 0.1386 | 1.7469 | 0.7571 | 2.3876 |
| KBr  | 0.1981 | 1.2084 | 0.0107 | 0.4617 | 0.6292 | 1.4892 |
| RbBr | 0.2894 | 1.1865 | 0.0238 | 0.6931 | 0.6423 | 1.6312 |
| LiI  | -      | -      | -      | -      | -      | -      |
| NaI  | 0.1472 | 1.1212 | 0.0078 | 0.4162 | 0.6973 | 1.6028 |
| KI   | -      | -      | -      | -      | -      | -      |
| RbI  | -      | -      | -      | -      | -      | -      |

[j] $E_{RI}=2KQ_0^2$ is a lattice reorganization energy. [k] $E_{CI} = E_{BI} + \frac{1}{2}E_{\alpha\beta}$ is a crossover energy. All these quantities are inherent to the off-center process.

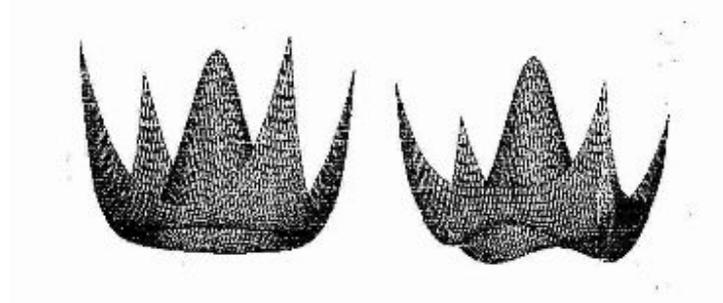

Figure 1:
Relief of the lower adiabatic potential energy surface (APES) as calculated using equation (14) with parameters from the KCl data in Tables I and II. For simplicity, one of the configurational coordinates has been set nil. Part (left) shows the APES resulting from first-order electron-mode coupling only, while part (right) obtains incorporating the third-order coupling terms as well. The comparison manifests that rotational barriers along the Sombrero brim appear as a third-order effect.